\documentclass[english]{article}
\usepackage[utf8]{inputenc}
\usepackage[english]{babel}
\usepackage[table,xcdraw]{xcolor}
\usepackage[normalem]{ulem}

\usepackage{geometry}
\usepackage{float}
\usepackage{amsmath}
\usepackage{graphicx}
\usepackage{natbib}
\usepackage{lipsum}
\floatstyle{ruled}
\newfloat{algorithm}{tbp}{loa}
\providecommand{\algorithmname}{Algorithm}
\floatname{algorithm}{\protect\algorithmname}

\newcommand{\lyxaddress}[1]{
	\par {\raggedright #1
	\vspace{1.4em}
	\noindent\par}
}

\begin{document}

\title{Particle-In-Cell Simulations of the Cassini Spacecraft's Interaction with Saturn's Ionosphere during the Grand Finale}


\author{Zeqi Zhang$^1$, Ravindra T. Desai$^{1}$, Yohei Miyake$^2$, Hideyuki Usui$^2$,  Oleg Shebanits$^{1,3}$}

\date{}
\maketitle

\vspace{-2em}
\lyxaddress{\begin{center}$^1$Blackett Laboratory, Imperial College London, London, UK \par\end{center}}
\vspace{-2em}
\lyxaddress{\begin{center}
$^2$Education Center on Computational Science and Engineering, Kobe University, Kobe, Japan
\par\end{center}}
\vspace{-2em}
\lyxaddress{\begin{center}
$^3$Swedish Institute of Space Physics, Uppsala, Sweden
\par\end{center}}
\vspace{-2em}

\begin{abstract}
 
A surprising and unexpected phenomenon observed during Cassini's Grand Finale was the spacecraft charging to positive potentials in Saturn's ionosphere. Here, the ionospheric plasma was depleted of free electrons with negatively charged ions and dust accumulating up to over 95 \% of the negative charge density. To further understand the spacecraft-plasma interaction, we perform a three dimensional Particle-In-Cell study of a model Cassini spacecraft immersed in plasma representative of Saturn's ionosphere. The simulations reveal complex interaction features such as electron wings and a highly structured wake containing spacecraft-scale vortices. The results show how a large negative ion concentration combined with a large negative to positive ion mass ratio is able to drive the spacecraft to the observed positive potentials. Despite the high electron depletions, the electron properties are found as a significant controlling factor for the spacecraft potential together with the magnetic field orientation which induces a potential gradient directed across Cassini's asymmetric body. This study reveals the global spacecraft interaction experienced by Cassini during the Grand Finale and how this is influenced by the unexpected negative ion and dust populations.

\end{abstract}

Key words:  space vehicles -- plasmas -- planets and satellites: individual: Saturn -- planets and satellites: atmospheres -- planets and
satellites: composition

\vspace{2em}
Accepted 2021 March 8. Received 2021 March 5; in original form 2021 January 22

\section{Introduction}

 Spacecraft immersed within a plasma will become electrically charged due to currents of incident electron and ion species \citep{Whipple81,Garrett81}. When the net current induced by these species are not zero, net charge is accumulated and there is a potential difference between the spacecraft and the surrounding plasma. Since the plasma species have different charges, each current is either decreased or increased by the change of the spacecraft potential, which continues until an equilibrium potential is reached, where the net current of all plasma species sum up to zero. 
Understanding how the spacecraft potential is affected by its environment is important for interpretting the surrounding plasma conditions and on-board plasma measurements which can be significantly affected by the potential difference. 

Near a given spacecraft surface a sheath boundary layer screens the potential of the surface over distances of the order of the Debye length \citep{Robertson13} and, for sufficiently high relative velocities between the spacecraft and surrounding plasma, a wake of depleted plasma is produced behind the obstacle \citep{Alpert66,Ludwig12,Miloch14}. 
As the spacecraft absorbs incident plasma, a shock is unable to form upstream as would for an obstacle able to withstand the incoming flow. If the relative velocity exceeds the Alfv\'enic or acoustic velocities, the associated discontinuities will instead trail downstream and form an approximately conical wake region of disturbed flow commonly referred to as a Mach cone \citep{Willis11}.
Significant asymmetries are introduced into the spacecraft-plasma interaction by the presence of a strong ambient magnetic field which generates a convective electric field in the spacecraft frame \citep{Marklund94,Pecseli12} and modifies the structure of the wake \citep{Darian17,Usui19}. The high velocities of the significantly lighter electrons compared with ions mean that a spacecraft immersed within a typical space plasma will charge to negative potentials \citep{Spitzer41} although processes such as photoelectron and secondary electron emission, can shift the potential to positive values \citep{Roussel04,Engwall06,Miloch09,Yaroshenko11}.

When passing through Saturn's ionosphere, the Cassini spacecraft's floating potential was, surprisingly, observed as positive on all encounters below 3000 km altitude \citep{Morooka19}. The presence of negative ions and dust grains in Saturn's ionosphere is evident from the LP measurements \citep{Morooka19}, which showed significant concentrations up to over 95\% of the negative charge density, at altitudes of 3200 km down to the closest measurement at $\approx$1600km. These appear an intrinsic part of the giant planet's ionosphere and distinct from electrons depletions associated with Saturn's main rings \citep[e.g.][]{Farrell18} and transient negative ion populations observed near Saturn's icy satellites \citep{Coates10,Desai18,Nordheim20}.
Unfortunately, it was not possible to obtain a mass distribution of these negative ions or dust grains due to Cassini's plasma spectrometers being offline and the Grand Finale plasma datasets therefore lack crucial pieces of information.  

A body immersed in a plasma with large quantities of negative ions can attain a positive potential due to the reduced electron currents, as was shown for a dust grain by \citet{Kim06} using orbital motion limit theory, and provides some indication for what might be occurring with Cassini. Similar electron depletions of up to $\approx$96 \% were, however, observed by Cassini within Titan's ionosphere where the spacecraft potential was consistently negative \citep{Wahlund05,Crary09,Shebanits16,Desai17a}. Significant unknowns therefore remain regarding how spacecraft interact with these outer solar system plasmas. 

Plasmas with a significant negative ion content (ion-ion or dusty plasmas) possess very different characteristics to typical electron-ion plasmas. The reduced mobility of the heavier negative charge carriers alters the electric field screening phenomenon and increases Debye length scales. Plasma conductivities can reverse \citep{Muralikrishna06,Shebanits20} and the plasma can host a variety of altered instabilities and wave phenomenon \citep{Shukla02,Desai17b}. The interaction of Cassini with Saturn's ionosphere therefore represents a class of physics quite different to the classic view of spacecraft charging within space plasmas.

In this article, we describe a self-consistent three dimensional Particle-In-Cell (PIC) study where a model Cassini spacecraft is simulated immersed within plasmas representative of Saturn's ionosphere as observed during the Grand Finale. Section \ref{method} describes the simulation approach, how Cassini is modelled, and the available measurements of Saturn's ionospheric plasma. Section \ref{Results} then applies these simulations to Saturn's ionosphere with regards to specific and general parameterisations and conducts a parametric survey to assess the sensitivity of the results to the measured and inferred plasma properties. Section \ref{summary} concludes with a summary of the key findings.

\section{Simulation Development}
\label{method}

\subsection{EMSES} 
\label{EMSES}

The three dimensional Particle-In-Cell simulation code Electro-Magnetic Spacecraft Environment Simulation (EMSES) has been developed for the self-consistent analysis of spacecraft-plasma interactions on either an electromagnetic or electrostatic basis \citep{Miyake09}. The electrostatic version of EMSES is utilised as the typical Alfv\'en velocities in Saturn's ionosphere are significantly greater than the spacecraft velocity and in this regime Alfv\'enic perturbations are assumed to form only a small contribution to the plasma currents \citep{Rehman14}. 

Within the PIC approximation \citep{Hockney81,Birdsall85}, individual particles are represented by large quantities of ``super-particles'' which are integrated through the same equations of motion as a real particle \citep{Boris70}. 
The plasma is defined to consist of an arbitrary number of plasma species in a drifting Maxwellian velocity distribution to represent in-flowing plasma. Each species has mass and charge normalized to the proton scale with a real ion-to-electron mass ratio. In this study a negatively charged ion component is included, which is required to study the target plasma environments. 
The positions of the super-particles are tracked continuously in the simulation box, but the electric and the magnetic fields are assigned and updated only on the grid points based on the \citet{Yee1966} algorithm,  and interpolated onto the super-particles' positions during calculation. The charge density profile at each timestep is used to solve Poisson's equation for the electrostatic potential with Dirichlet boundary conditions.

The simulations are run in the spacecraft frame and consists of a three dimensional box with inflow and outflow boundary conditions along a specified direction of plasma flow, and periodic boundary conditions orthogonal to these.
Within the domain the spacecraft is considered as a perfect conductor with separate boundary treatments for both longitudinal and transverse electric fields. The spacecraft surface can accumulate charge caused by impinging super-particles, and redistributes this to ensure the spacecraft is equipotential using the capacity matrix method \citep{Hockney81,Miyake09}. 

\begin{figure*}[ht]
\centering
\hspace{-3em}
\includegraphics[width=0.9\textwidth]{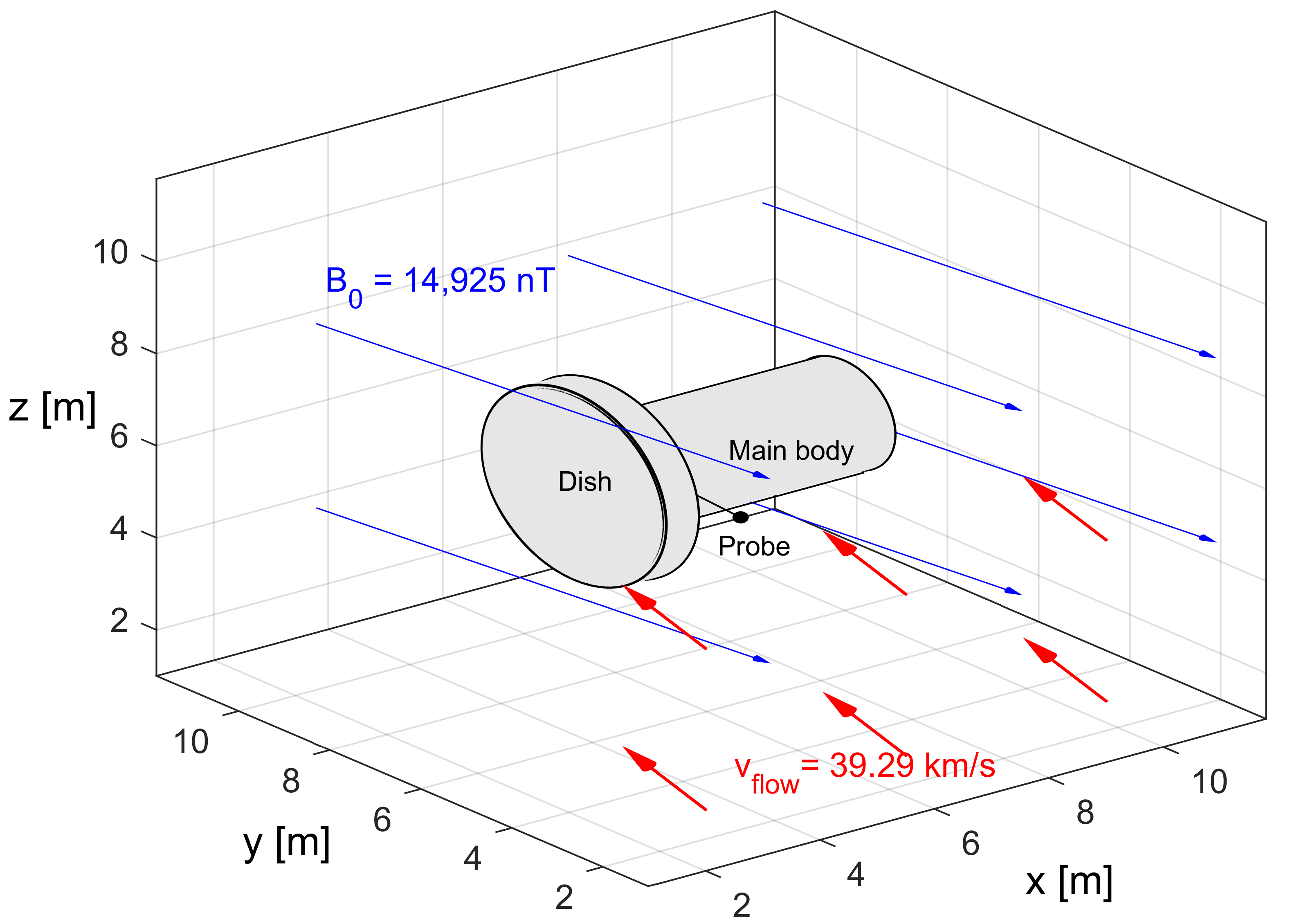}
\caption{Simulation configuration for Cassini during Grand Finale Rev 292 ingress at 2500 km Saturn altitude. The X, Y and Z axes correspond to the Z$_{s/c}$, X$_{s/c}$, and Y$_{s/c}$ axes of the Cassini spacecraft attitude coordinate system, respectively, and the precise simulation and plasma parameters are provided in Table \ref{table}.
\label{cassini}}
\end{figure*}

\subsection{Application to Cassini}
\label{model}

\subsubsection{Main Body}
\label{body}
The Cassini spacecraft is a three-axis stabilised spacecraft of approximately six metres in length. Previous spacecraft charging simulations have considered Cassini in two dimensions \citep{Olson10} and as a cylinder in three dimensions during the 2004 Saturn orbit insertion \citep{Yaroshenko11}. In this study we model Cassini's non-uniform shape encompassed within the simulation domain of 12.8$\times$12.8$\times$12.8 m$^3$, across 128$\times$128$\times$128 grid cells. Cassini is approximated using three structures: a large thin cylinder representing the antenna dish with an approximated diameter of 4 metres and width of 0.55 metres, and a longer cylinder representing the main body with an approximated diameter of 2.2 metres and length of 3.8 metres. These are separated by a 0.55 metres gap between them, as per the real Cassini spacecraft. Figure \ref{cassini} shows a schematic of the simulation geometry where the X, Y and Z axes correspond to the Z$_{s/c}$, X$_{s/c}$, and Y$_{s/c}$ axes of the Cassini spacecraft attitude coordinate system, respectively: where X is the direction of main thrusters, Y is opposite to the Langmuir probe direction and Z completes the right-hand set.

The dish and the main body of the spacecraft are considered as a single perfect conductor, as Cassini was designed. Although this approximation does not account for the curvature of the antenna dish, or other instruments attached onto the body, the most important factors for spacecraft charging are the surface area of structures larger than the Debye length scales as well as the ram profile for a fast moving object.
As far as the net surface current of the spacecraft and its potential evolution are concerned, this approximation is judged to reasonably approximate dynamics associated with Cassini's asymmetric shape interacting with Saturn's ionosphere.

\subsubsection{Langmuir Probe}
\label{lprobe}
The Langmuir Probe (LP) is represented by a small sphere at the side of Cassini, the precise extent of which is defined subgrid.
 The LP is held at a bias to the main spacecraft, the precise voltage difference specified according to which point in the LP voltage sweep between --4 V and 4 V that is simulated. The time required for the currents to equilibriate is small compared to the sweep time scale of 0.5 s \citep{Gurnett04}, and a fixed bias within the sweep is chosen. The whole Cassini model is floating within the plasma environment and can therefore become charged relative to it but the voltage between Cassini and the probe remains fixed. 
 
For a Langmuir Probe at a net negative potential within an electron-ion plasma, the total external plasma currents can be considered to derive purely from the positive ion current. This is because the lighter electrons are repelled and therefore their current contribution can be neglected. Similarly, at a net positive potential the external plasma currents is saturated by the faster moving electrons and can be considered to purely consist of electrons.
In the presence of large negative ions, however, the situation is different. When the LP is negatively biased, the large negative ions and dust grains are still collected due to their high inertia (mass). At positive biases the current will be a combination of electrons and negative ions but the significantly smaller electrons still dominate the total current. 
The simulations are therefore run with the Langmuir Probe biased at a negative potential of --3 V to ensure no electrons are collected as the bias potential is higher than the electron thermal energy. The self-consistent addition of the negative ion component therefore allows us to constrain their contribution to the Langmuir Probe currents and to the overall potential accumulated by the spacecraft.

\subsection{Saturn's Ionospheric Plasma}
\label{Measurements}
Within Saturn's ionosphere the total current onto the spacecraft is a function of the positive ions, electrons and negative ions, charged dust, secondary electron emission and photoelectron emission. The photoelectron current at Saturn's orbit are several orders of  magnitude lower than the ion and electron currents \citep{Holmberg17,Shebanits17}, and the secondary electron emission currents are assumed to be of a similar magnitude to this \citep{Morooka19}. They are therefore not included in this analysis.  
The simulated currents can therefore be expressed as
\begin{equation}
    I_{total} =
    I_{electron} +  I_{ion^+} + 
    I_{ion^-}, 
    \label{eqcurrents}
\end{equation}
where  I$_{ion^+}$ represents the positive ions and dust and I$_{ion^-}$, represents the negative ions and dust. While multiple positive and negative plasma components can be included the lack of information on the mass distribution function leads us to use a mean mass to represent these respective components.

\subsubsection{Cassini Observations}
\label{observations}

The measured composition of Saturn's ionosphere provides the inputs to the simulations which produce the currents in Equation \ref{eqcurrents}. Post--2012 CAPS was, unfortunately, turned off and the mass distribution on the ambient plasma in Saturn's ionosphere was therefore unknown. Cassini's Ion and Neutral Mass Spectrometer was able to provide some information on positive ion composition but this was limited up to low masses of several amu due to the high spacecraft velocity \citep{Waite18}. The Langmuir Probe was, however, able to differentiate between the bulk ion, electron and negative ion currents \citep{Wahlund18,Hadid19,Shebanits20} and therefore provide estimates of their densities \citep{Morooka19}.
 The electron density was determined from the positive bias side of the Langmuir Probe sweep \citep{Morooka19}. The positive ion density could, however, only be constrained as a lower limit due to the unknown mass distribution \citep{Shebanits13,Morooka19}. A lower limit on the negative ion density was therefore also available by assuming quasi-neutrality. 

\begin{table}[ht]
\caption{Environmental and System Simulations Parameters}
\centering
\begin{tabular}{|
>{\columncolor[HTML]{CBCEFB}}l 
>{\columncolor[HTML]{CBCEFB}}c |}
\hline
\multicolumn{2}{|c|}{\cellcolor[HTML]{CBCEFB}\textbf{Environmental Parameters}} \\                   
Plasma ion density, $n_0$                            & 854  cm$^{-3}$ \\
Negative ion concentration, $n_{ni}$                       & 40.2 \%                     \\
Ion mass, $m_i$                                  & 1.4 amu                     \\
Negative ion mass, $m^{-}_{i}$                        & 3.0 amu                     \\
Electron temperature, $T_e$                      & 0.093 eV                      \\
Ion temperature, $T_i$                           & 0.093 eV                      \\
Negative ion temperature, $T^-_{i}$                & 0.093 eV                      \\
Magnetic field, $\vec{B}$                            & [1.48$\hat{x}$, --14.8$\hat{y}$, 1.24$\hat{z}$] $\mu$T                     \\
Flow velocity, $\vec{v}_{flow}$                             &  [--0.189$\hat{x}$, --37.3$\hat{y}$ --12.2$\hat{z}$] km s$^{-1}$                   \\

Alfv\'en speed, $v_{A}$             & 11,149 km s$^{-1}$                   \\
Ion acoustic speed, $v_{S}$             & 2.56 km s$^{-1}$                    \\
Debye length, $\lambda_D$                         & 10.05 cm                    \\
Electron gyroperiod, $\tau_{ge}$                   & 2.38 $\mu$s                     \\
Electron plasma period, $\tau_{pe}$                  & 3.81 $\mu$s                    \\
Ion gyroperiod, $\tau_{gi}$                        & 6.12 ms                     \\
Ion plasma period, $\tau_{pi}$                       & 0.193 ms                    \\
Negative ion gyroperiod $\tau^-_{gi}$ & 13.1 ms                     \\
Negative ion plasma period, $\tau^-_{pi}$             & 0.283 ms                    \\
\multicolumn{2}{|c|}{\cellcolor[HTML]{CBCEFB}\textbf{System Parameters}}                               \\
Grid width, $\Delta$r                           & 10 cm                       \\
Time step, $\Delta$t                            & 0.033 $\mu$s                   \\
Simulation time, $t$                            & 0.67 ms                     \\
Particles per cell                        & 20                          \\
Domain size                                   & 12.8$^3$ m$^3$  \\
Probe Bias, $\phi_{LP}$                             & --3 V                        \\ \hline
\end{tabular}
\label{table}
\end{table}
 
 The electron density was observed to increase with decreasing altitude as Cassini sampled denser regions of Saturns ionosphere \citep{Persoon19,Morooka19}. The estimated negative ion density was also observed to increase and at a greater rate than the electrons. As electrons are lost to the negative ions/dust grains, the electron depletion also increased with decreasing altitude, and reached an estimated lower bound of $94$ \% \citep{Morooka19}. The electron depletion is traditionally quoted as a ratio of negative to positive ion densities such that a 94 \% depletion corresponds to 6 \% of the total negatively charged species being electrons, with the rest being negative ions. The Langmuir Probe determined a mean electron temperature of $\approx$0.1 eV and, based upon current balance, an estimate of the minimum mean positive ion mass of $\approx$5 amu at the deepest point sampled \citep{Morooka19}. The electron temperature was used as a conservative upper limit on the ion temperatures and reducing the ion temperatures below this was later found not to have a  significant influence on the simulation results. Within Saturn's ionosphere the spacecraft potential varied between --1.25 and +0.75 V with a positive floating potential on every encounter below 3000 km.

\subsubsection{Simulation Inputs}
\label{inputs}

Rev 292 is selected as a representative flyby with sufficient data from which input parameters are derived. In the first instance, Cassini is simulated at an altitude of 2500 km during ingress where the LP observations revealed an estimated plasma density of 854 cm$^{-3}$, temperature of 0.093 $eV$, and the lower bound on the estimated electron depletion was 40 \%. A higher electron density decreases the Debye length which is the primary constraint on the simulation grid size. This altitude therefore provided an optimal balance between computational load and studying regions of interest with large negative ion concentrations. The magnetic field was measured to be 14,925 nT which results in the electrons being highly magnetised with gyroradii of the order of centimetres and therefore significantly less than the spacecraft size. The positive and negative ions are only weakly magnetised, with gyroradii of tens to hundreds of metres, and therefore much larger than the spacecraft and the simulation domain. 
Collision frequencies in Saturn's ionosphere are much less than a herz \citep{Shebanits20} and the associated period is significantly greater than the total simulation time of less than a millisecond. During Rev 292 the spacecraft potential was --0.12 V at this altitude, and varied between --0.75 and +0.65 V. The precise simulations input parameters are provided in Table \ref{table}.

The relative orientation of the spacecraft to the plasma and magnetic field varies very slowly throughout a flyby and is different for different flybys. 
The simulations described herein do not attempt a produce a precise reproduction of the Cassini measurements, instead attempting to understand the key features of the plasma interaction and associated charge accumulated. 
\section{Results}
\label{Results}

\subsection{Global Interaction}
\label{global}

\begin{figure*}[ht]
\hspace{-2em}
\includegraphics[width=1.1\textwidth]{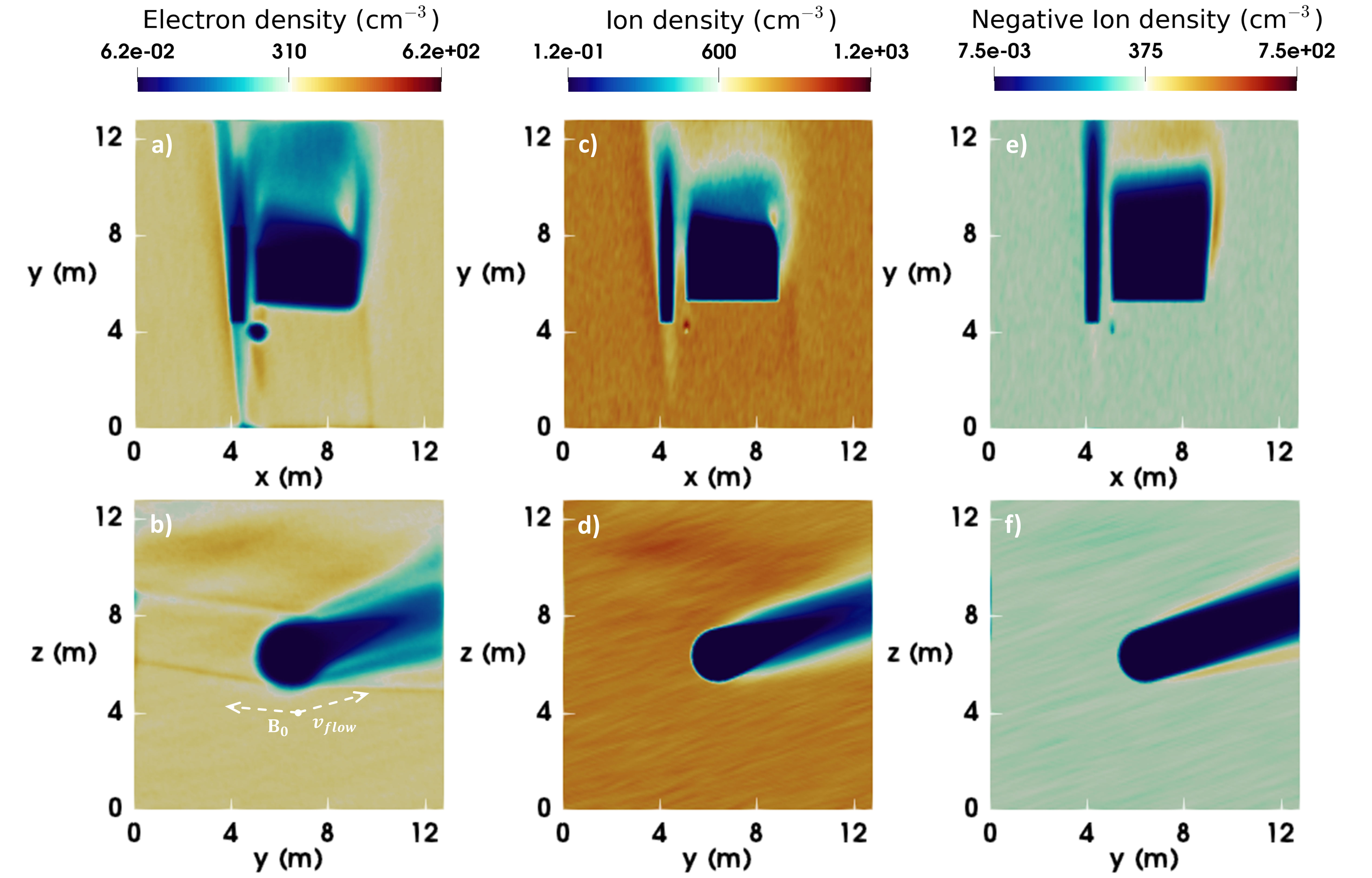}
\caption{Two dimensional slices of the electron (a--b), ion (c--d) and negative ion density (e--f) for simulation parameters outlined in Table \ref{table} and described in Section \ref{global}. The magnetic field is oriented along 1.48$\hat{x}$, --14.8$\hat{y}$, 1.24$\hat{z}$ and a schematic of the simulation geometry is displayed in Figure \ref{cassini}.}.
\label{global1}
\end{figure*}

The simulated interaction between Cassini and Saturn's ionosphere is shown at Figure~\ref{global1}. The left-hand (a--b), centre (c--d) and right-hand (e--f) panels show electron, ion and negative ion densities respectively, with the upper (a, c, e) and lower panels (b, d, f) respectively showing x--y and y--z slices through the simulation. The results are displayed at the end of the simulation run. The two cylinders representing Cassini's antenna dish and main body are apparent as distinct regions absent of plasma. The plasma velocity is arriving predominantly along the y-direction at an angle of $\approx 168 ^{\circ}$ degrees to the near-oppositely directed magnetic field. The Langmuir Probe is pointed in the spacecraft ram direction and therefore directly samples the incoming plasma flow. The probe is most visible in Figure~\ref{global1}(a) at x=4 and y=4 as a local depletion in the electron density due to the negative bias of --3 V. 

The high relative velocity of the plasma interaction produces an extended wake of depleted plasma.
The length of the wake is greater for positive and negative ions due to their larger inertia resulting in slower refilling. When looking at a slice through Cassini's main body, panels (b, d, f), the wake exhibits the characteristic structure associated with supersonic flow around a cylinder. The slice in the x-y plane, however, in panels (a, c, e) reveals the wake as non-uniform and highly structured behind the antenna dish and the gap between the antenna dish and the main body

The electron density reveals wing-like structures of enhanced and then depleted density attached to the spacecraft in the y-z plane in panel (b). Similar wing-like structures have been identified at moving bodies produced by Alfv\'en and whistler waves \citep{Drell65,Neuebauer70,Stenzel89}, which propagate at characteristic velocities associated with their relative speed to the spacecraft such that they advect downstream. Electron-wing structures, similar to these reported herein, have been identified as consisting of propagating Langmuir waves \citep{Miyake20} produced by electrons reflected from a negatively charged spacecraft which are then guided by the magnetic field lines. Calculation of the Langmuir Probe group velocity, c$_L$=$\sqrt{3 k_B T_e m_e }\cong$ 1,200 km s$^{-1}$, and the angle to the magnetic field, $\theta = \arctan({v_{flow}/c_L}) \approx 2^{\circ}$, confirms this mode propagates at small angles to the magnetic field in the spacecraft frame, as can be seen in Figure \ref{global1}.  Figure \ref{global1}a shows these wings structures striking the inflow boundary condition and modifying the electron density. Figure \ref{global1}b however reveals that the effects of this are shifted towards positive z-regions and  mostly miss the spacecraft interaction and therefore have a negligible numerical impact. The electron-wings propagate upstream of Cassini and, for flybys where the magnetic field is more closely aligned with the spacecraft velocity, they may therefore have influenced the properties of the assumed pristine plasma ahead of Cassini.

Different density distributions are also visible for the different species and the electrons particularly show enhanced spatial variations around the spacecraft. To either side behind the antenna dish in Figure \ref{global1}a two distinct regions of depleted flows are visible which appears to produce a vortex-type structure in the x-z plane. This is analysed further in Section \ref{Gradient}.

\begin{figure*}[ht]
\centering
\includegraphics[width=1\textwidth]{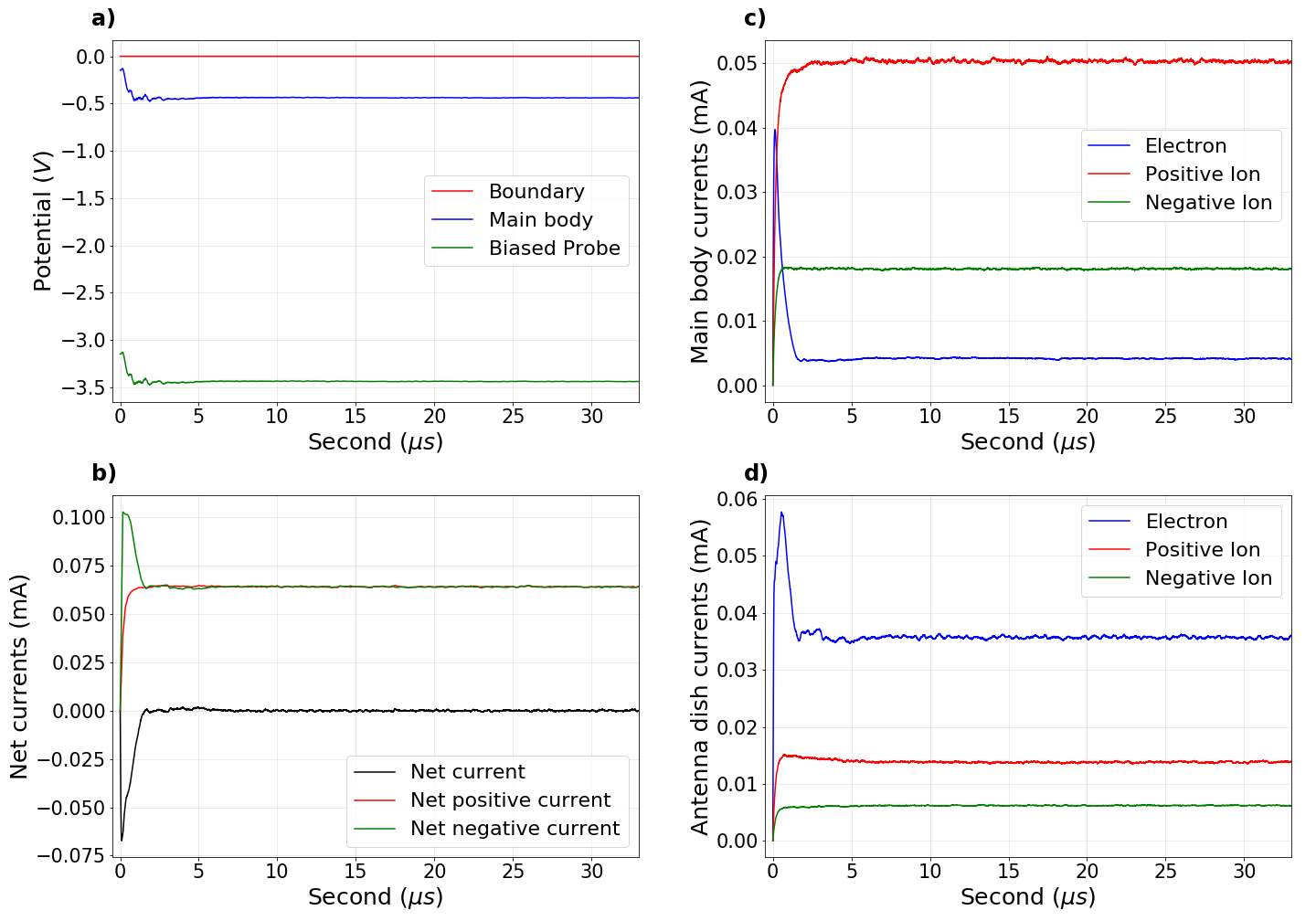}
\caption{Currents onto the spacecraft within Saturn simulations. (a) shows the spacecraft potential over time, (b) shows the current onto the spacecraft over time, (c) shows the current onto the dish over time, and (d) shows the current decomposed into the positive and negative parts.
\label{currents}}
\end{figure*}

\subsection{Spacecraft Potential}
\label{spacecraftpotential}

In this Section we compare the simulations results to Cassini Langmuir Probe measurements of the spacecraft potential and plasma currents during ingress on Rev 292. Figure \ref{currents} shows the evolution of the spacecraft potential and plasma currents through this simulation run for input parameters derived for when Cassini was at a Saturn altitude of 2500 km in the northern hemisphere. 

The net spacecraft potential is shown in panel (a) and incident plasma currents in panel (b), both of which start near zero. As the spacecraft charges, the main spacecraft body and the probe remain at a fixed bias relative to one another of 3 V. The net current becomes strongly negative before returning to zero when an equilibrium is reached after $\approx$1 $\mu$s. The spacecraft has at this point accumulated a negative floating potential of just under --0.49 V with the probe biased at --3.49 V. The simulation is, however, run for significantly longer to ensure steady state. The simulated potential is --0.42 V, and 0.3 V more negative than the observed potential of --0.12 V. The sensitivity of the spacecraft potential to variations of this magnitude are analysed within following Sections \ref{survey}.

\begin{figure*}[ht]
\centering
\includegraphics[width=0.8\textwidth]{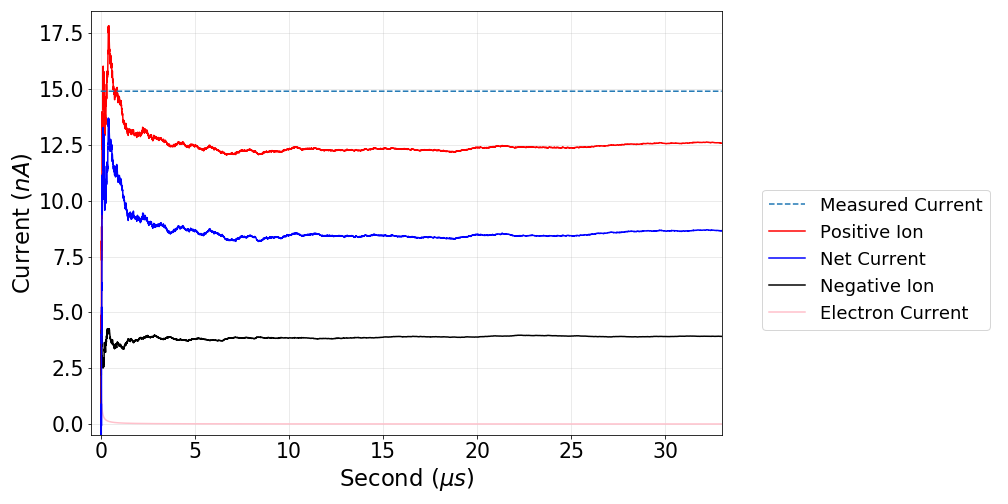}
\caption{Currents onto the simulated Langmuir Probe decomposed into the positive and negative parts and compared to the Cassini Langmuir Probe readings during Rev 292 at 2500 km altitude ingress.
\label{probe}}
\end{figure*}

The current decomposition for each of the electrons, positive and negative ions onto each part of the spacecraft is shown in panels (c--d). The spike in the negative current in panel (b) corresponds to electron current which also peaks at the beginning of the simulation. The spacecraft surfaces accumulate the associated negative charge and increasingly repel electrons away, thereby reducing the electron current. These initial electron dynamics reflects how the smaller electrons are much more sensitive to the change in the potential of the spacecraft compared to the negative ions which, due to their larger momentum, are not as easily deflected. 
At the equilibrium potential, all three species of the plasma contribute comparable level of currents onto the spacecraft relative to each other.
This current balance is significantly different to within typical electron-ion plasmas
where the electrons constitute the majority of the incident current \citep[][and references therein]{Miyake20}.  Comparing between the current received from the dish and from the main body, the current composition and relative magnitudes are also different. This is due to the difference in their surface areas but also the deflection of the particle's trajectories by the ambient magnetic field resulting in increased quantities of ions and negative ions striking the sides of the dish despite the incident flow being approximately parallel to it. 

\subsection{Langmuir Probe Currents}
\label{probecurrents}

Figure \ref{probe} shows the currents onto the simulated Langmuir Probe. The positive ion current stabilises after $\approx$ 10 $\mu$s at $\approx$ 125 nA, slightly lower than the observed total current of $\approx$ 150 nA.  However, when summing the net current over all plasma species, the current is around 30 \% less. The initial analysis of the LP current was conducted assuming the total current derived from positive ions and that the negative current was negligible.
As a result of the simulation it is now clear that the negative ion current is not negligible and can be approximated as an almost constant current, regardless of the potential difference, due to their inertia. These results highlight that while the electron density can be accurately determined, the total plasma density cannot due to the ion and negative ion components. Therefore, by scaling the plasma concentration the net currents onto the probe will increase. There could also be other effects at play such as from the impact and break-up of large dust particles onto the probe and spacecraft which are not included in these simulations.

\subsection{Parametric Survey}
\label{survey}

To further understand the sensitivity of the results to the unknown or inferred parameters, we carry out a parametric study and systematically varied the following parameters; the electron depletion, the ion and negative ion masses, the electron temperature and also the ion and negative ion temperatures which were previously assumed to be in thermal equilibrium with the electrons. The total plasma density, however, remains unchanged and the densities are scaled within this. The positive ion mass is now set at 5 amu to represent the measured lower bound in the deep ionosphere \citep{Morooka19}.

\begin{figure*}[ht]
\centering
\includegraphics[width=1\textwidth]{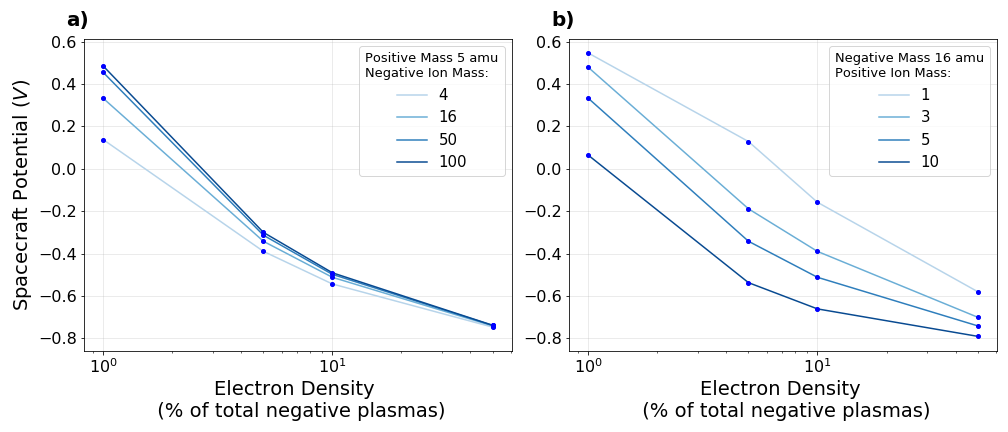}
\caption{Parametric scan of electron density depletion for a range of negative ion masses (a) and positive ion masses (b).}
\label{scanmass}
\end{figure*}

Figure \ref{scanmass}(a) shows the spacecraft potential as a function of electron depletion. The electron density is varied from 1 \% up to 50 \% of the total ion density, for negative ion masses of 4, 6, 50 and 100 amu. For all negative ion masses simulated, as the electron density tends to zero, the spacecraft potential becomes less negative and eventually reaches positive values. For larger negative ion masses the spacecraft become positive at slightly higher electron densities although this change becomes increasingly smaller with the difference between 50 and 100 amu being significantly smaller than that between 4 and 50 amu. 

The effect of a varying positive ion mass on the spacecraft potential is shown in Figure \ref{scanmass}b where the negative ion mass is held constant at 16 amu and the spacecraft similarly tends to positive potentials for increased electron depletions. Smaller positive ion masses produced positive potentials at lower masses than larger ones and this change also becomes smaller for the larger masses. The spacecraft potential also appears more sensitive to the positive ion mass than to the negative ion mass. 
These trends are explained by the variation in the relative mobilities of the positive and negative charge carriers. The production of a positive potential is due to the positive ions actually becoming the most mobile charge carriers which result in the spacecraft accumulating a net positive current. \citet{Kim06} indeed predict that a body immersed in an electron-depleted negative ion plasma can gain a positive potential when the electron density reaches just a few percent of the total plasma density. Figure \ref{scanmass}a demonstrates this is possible for the Cassini spacecraft in Saturn's ionosphere.


\begin{figure*}[ht]
\centering
\includegraphics[width=1\textwidth]{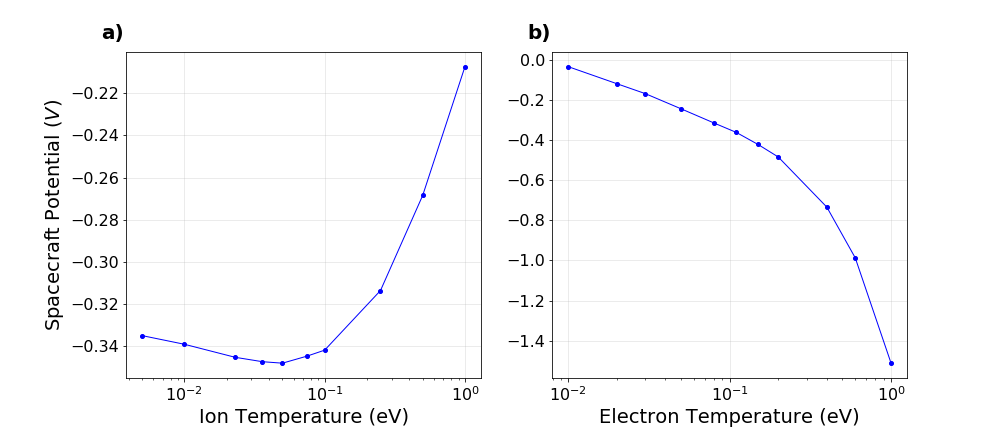}
\caption{Parametric scan of ion (a) and electron (b) temperatures.}
\label{temperaturescan}
\end{figure*}

To study the sensitivity to the electron and ion temperatures, the negative ion mass is held at 16 amu and positive ion mass at 5 amu but the depletion rate is now set at 95 \% to represent the parameter regime close to where Cassini gains a positive potential. The ion temperature variation, shown in Figure \ref{temperaturescan}a, reveals that the spacecraft potential increases as the temperature increases from the measured mean temperature of $\approx$0.1 eV, reaching --0.21 V when T$_{i}$=1 eV. For temperatures below the measured 0.1 eV, the spacecraft potential initially follows this trend but then, surprisingly, increases again when T$_{i} \leq$0.5 eV. 
This slight effect is attributed to the ions becoming increasingly magnetised as the mean ion gyroradius decreases until it is of a comparable size to the spacecraft. The electron temperature variation, shown in Figure \ref{temperaturescan}b, has a much larger impact on the spacecraft potential due the electron's high relative mobility. The potential consequently varies from -0.05 V to -1.5 V, despite the electrons constituting just a few percent of the total plasma density. 

\subsection{Potential gradient}
\label{Gradient}

The relative mobility of the positive and negative ions explains the positive potential. However, in Figure \ref{scanmass}a, the spacecraft still charges to a positive potential when using negative ions lighter than positive ions (4 and 5 amu respectively), so that the negatively charged particles remain the more mobile charge carriers. This therefore appear to contradicts this explanation. Beyond the variation in the plasma parameters examined in Section \ref{survey}, the effect of the ambient magnetic field remains to be evaluated. 

\begin{figure*}[h]
\centering
\includegraphics[width=0.7\textwidth]{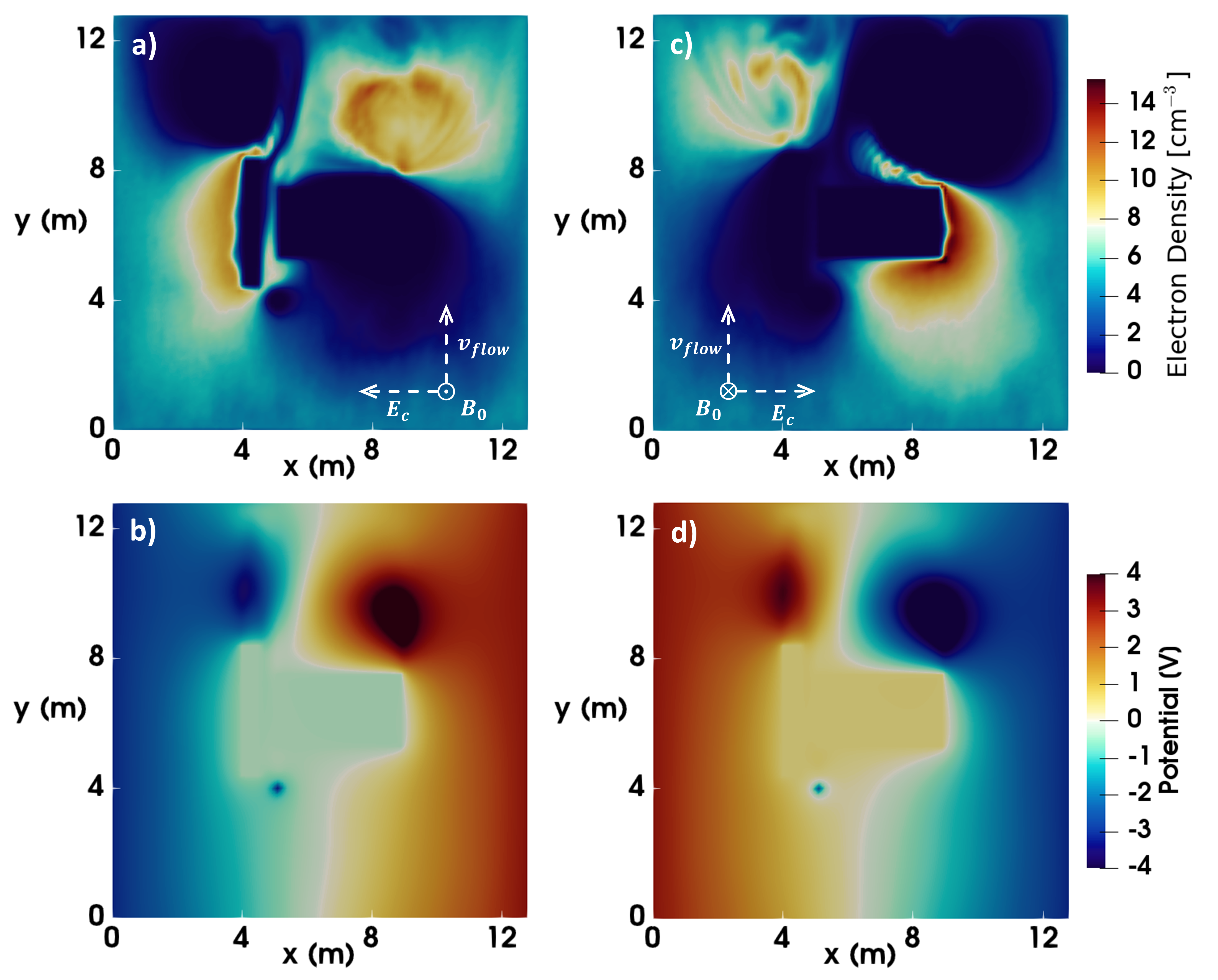}
\caption{Electron density (a \& c) and potential (b \& d) of Cassini immersed in a plasma with a negative ion mass of 4 amu and positive ion mass of 5 amu. The left-hand panels (a \& b) show the case where the magnetic field is oriented along B$_z$ which produces a spacecraft potential of --0.42 V whereas the right-hand panels (c \& d) show the case with the magnetic field reversed where the spacecraft potential is +0.38 V.
\label{potential}}
\end{figure*}

Within the next simulations the magnetic field direction is oriented perpendicularly to the plasma flow direction, instead along the +z and --z axis. The influence of the magnetic field is felt in the generation of the convective electric field, E$_c$, in the spacecraft frame, namely,
\begin{equation}
    \boldsymbol{E_c} = \boldsymbol{-v}_{flow} \times \boldsymbol{B_0},
\end{equation}
which produces a potential gradient along the main axis of the spacecraft \citep{Pecseli12}. 
Figure \ref{potential} shows the electron density and potential within the two simulations with the two different field orientation. The electrons can be seen to be accumulating on one sides of the spacecraft and the potential gradient across the spacecraft interaction is clearly visible. The potential distribution controls where the electrons impact the spacecraft and alters the effective spacecraft cross section due to Cassni's asymmetric shape. The spacecraft potential is consequently +0.38 V in the initial magnetic field configuration but --0.42 V with the field reversed. Further tests run without the presence of a magnetic field (not shown) also show that the spacecraft is unable to attain a positive potential when the positive ions are less massive than the negative ions.

The perpendicular magnetic field enhances the magnetisation of the plasma interaction. The differential electron flows around Cassini can be seen to produce highly structured vortices at a scale that is comparable to Cassini itself. It is interesting to note that electrostatic spacecraft-scale modes have been identified in Saturn's deep ionosphere which are thought to be modulated by Cassini's presence \citep{Sulaiman17}.

\section{Conclusions}
\label{summary}

This study simulated the spacecraft-plasma interaction experienced by the Cassini within Saturn's ionosphere during the Grand Finale using a self-consistent three-dimensional PIC code. The asymmetric spacecraft shape was taken into account and a Langmuir Probe was also modelled which enabled preliminary comparisons between the simulated and observed currents.

The global interaction of the Cassini spacecraft was found to be strongly influenced by its asymmetric shape. Differential flows around Cassini's antenna produced a highly structured wake and spacecraft-scale electron vortices behind the spacecraft. Electron-wings associated with propagating Langmuir waves were also identified around the spacecraft which produced a sharp enhancements to the in-flowing electrons. Similar electron-wings were previously identified in simulations of spacecraft moving through the polar regions of the Earth's ionosphere \citep{Miyake20}  and these simulations indicate this phenomena occurred at Cassini too.

An electron density of 40 \% of the total negative charge density was initially used in this simulations, which was determined to be a lower bound by Cassini's Langmuir Probe at 2500 km Saturn altitude during Rev 292 \citep{Morooka19}. The spacecraft potential was found to be negative in the same range as that observed at --0.42 V compared with the observed --0.12 V. The simulations indicated that in Saturn's ionosphere, the ion currents onto the spacecraft and Langmuir Probe were comparable to those from the electrons, a situation unique to plasmas depleted of electrons. The simulated currents onto the Langmuir probe also revealed that the total currents were in the same range, although slightly lower, than those measured. This is consistent with the 40 \% electron depletion being near the lower bound and indicates that the actual positive and negative ion density may have been higher.

Following these initial simulations a parametric study was carried out to examine how the floating potential is affected by the ambient conditions. To this end, we studied how the spacecraft potential changes with positive and negative ion mass and temperature as well as electron depletion rate.
As the electron depletion rate increased the spacecraft potential was observed to become less negative and even attained positive potentials of up to 0.55 V when just a few percent of electrons were contained within the plasma. Varying the plasma masses and temperatures had a similar effect in terms of charge mobility with the electron temperature found to have the largest influence.

The magnetic field orientation was also varied to examine how the induced electric field along Cassini's main axis affected the charge accumulated. 
The direction of the magnetic field, therefore electric field, produces a strong potential gradient in the potential of the plasma and results in oppositely charge currents preferentially collecting at different parts of the spacecraft. Given the large different in surface area across Cassini due to its large antenna dish, reversing the magnetic field actually shifted the spacecraft  through zero from an initial 0.38 V to negative values of a comparable magnitude of --0.42 V.

As well as providing an insight into the spacecraft plasma interactions experienced by Cassini during the Grand Finale, these simulations have provided an explanation for the positive potentials attained through considering classical charging theory and reversed charge mobility. Despite this, further effects such as secondary electron emission and the break-up of large dust or ice particles on the spacecraft, and an even more detailed Cassini model incorporating effects such as the curvature of the antenna dish, remain to be constrained. The study does indicate, however, that this approach presents a valuable method for understanding spacecraft charging in the outer solar system and can assist in interpretting in-situ measurements of these exotic environments. 

\section*{Acknowledgements}
 ZZ acknowledges an Undergraduate Summer Research Bursary from the Royal Astronomical Society. RTD acknowledges funding from NERC grant NE/P017347/1. YM and HU acknowledge grant 20K04041 from the Japan Society for the Promotion of Science: JSPS,
and support from the innovative High-Performance Computing Infrastructure (HPCI: hp200032) in Japan. OS acknowledges RS grant RP EA180014 and SNSA grant Dnr:195/20. This work has benefited from discussions with International Space Science Institute (ISSI) International Team 437. This work used the Imperial College High Performance Computing Service (doi: 10.14469/hpc/2232). 
\newline

\section*{Data Availability}
All simulation data presented in this study can be retrieved from the Zenodo open-access repository at https://doi.org/10.5281/zenodo.4592954. Cassini Langmuir Probe data is available from the NASA Planetary Data System archive.
\bibliography{EMSES_bib}




\end{document}